\documentclass[12pt]{iopart}

\usepackage{iopams}
\usepackage{setspace}

\usepackage{amsfonts}   
\usepackage{amssymb}
\usepackage{graphicx}   

\begin{document}
\title{Geometric entanglement in the Laughlin wave function}
\author{Jiang-Min Zhang$^{1,2,*}$, Yu Liu$^{3,4,\dag}$ }
\address{$^1$ Fujian Provincial Key Laboratory of Quantum Manipulation and New Energy Materials,
College of Physics and Energy, Fujian Normal University, Fuzhou, 350007, China}
\address{$^2$ Fujian Provincial Collaborative Innovation Center for Optoelectronic Semiconductors and Efficient Devices, Xiamen, 361005, China}
\address{$^3$ LCP, Institute of Applied Physics and Computational Mathematics,
Beijing, 100088, China}
\address{$^4$ Software Center for High Performance Numerical Simulation,
Chinese Academy of Engineering Physics, Beijing, 100088, China}

\ead{$^*$wdlang06@163.com, $^{\dag}$liu\_yu@iapcm.ac.cn}

\begin{abstract}
We study numerically the geometric entanglement in the Laughlin wave function, which is of great importance in condensed matter physics. The Slater determinant having the largest overlap with the Laughlin wave function is constructed by an iterative algorithm. The logarithm of the overlap, which is a geometric quantity, is then taken as a geometric measure of entanglement. It is found that the geometric entanglement is a linear function of the number of electrons to a good extent. This is especially  the case for the lowest Laughlin wave function, namely the one with filling factor of $1/3$. Surprisingly, the linear behavior extends well down to the smallest possible value of the electron number, namely, $ N= 2$. The constant term does \emph{not} agree with the expected topological entropy. In view of previous works, our result indicates that the relation between geometric entanglement and topological entropy is very subtle.
\end{abstract}



\submitto{\NJP}
\maketitle

\section{Introduction}

The Laughlin wave function \cite{laughlin} is a paradigm as a variational wave function \cite{nobel}.
Although simple in construction, its implication is rich. It not only explains the most robust fractional quantum Hall effects \cite{tsui}, but also embodies many exotic notions like fractionally charged excitations, fractional statistics, and composite-fermions/bosons \cite{yoshioka,Chakraborty}.

The simplicity and fertileness of the Laughlin wave function make it very intriguing, and significant efforts have been dedicated to its nature \cite{johnson}. Besides the conventional way of studying correlation functions, a more recent approach is studying the entanglement in the wave function \cite{ zeng, shi, masud}. In \cite{masud}, bipartite entanglement entropy in the Laughlin wave function has been studied both analytically and numerically. A noteworthy result is that, by studying the scaling behavior of the entanglement entropy, the topological entropy corresponding to the so-called ``total quantum dimension'' \cite{ kitaev,wen} is extracted, confirming the fact that the Laughlin wave function is a topologically ordered state.

However, a most natural way of quantifying the entanglement in the Laughlin wave function has not yet been explored so far. It is well known that unlike the integer quantum Hall effect, for the fractional quantum Hall effect, the Coulomb interaction between the electrons is essential. For a non-interacting fermionic system,  the multi-particle eigenstates are very simple---They are Slater determinants \cite{slater}, the simplest kind of fermionic wave function satisfying the antisymmetry condition. For a generic interacting fermionic system, the interaction mixes up the Slater determinants and a generic eigenstate $\Phi$ can no longer be written in the form of a Slater determinant \cite{interaction}, no matter how the single-particle orbitals are chosen. Specifically, as we shall prove below, the Laughlin wave function cannot be written as a single Slater determinant. Now, from the point of view of approximation theory \cite{timan}, a  natural question is, how close can it be approximated by a Slater determinant? This question leads to the notion of the optimal Slater approximation of a fermionic wave function \cite{zjm,zjm2}, namely, the Slater determinant which maximizes the overlap
\begin{eqnarray}
  \mathcal{O} &=& |\langle S | \Phi \rangle |^2,
\end{eqnarray}
where $S$ denotes a Slater determinant.

Denote the maximal value of $\mathcal{O}$ as $\mathcal{O}_{\rm{max}}$. It takes values from the interval $(0,1 ]$.  Apparently, the larger $\mathcal{O}_{\rm{max}}$ is, the closer the original wave function $\Phi$ is to a Slater determinant, and it is reasonable to say that the less the fermions are entangled with each other. In particular, if $\mathcal{O}_{\rm{max}} = 1$, the wave function is a Slater determinant and the entanglement is zero. Therefore, a measure of entanglement can be defined as
\begin{eqnarray}
  E_G (\Phi ) &=& - \ln ( \mathcal{O}_{\rm{max}} ) .
\end{eqnarray}
This is a geometric measure of entanglement in the sense that it involves only the notion of inner product between two states in a Hilbert space. It is a natural fermionic generalization of the geometric measure of entanglement for spin systems \cite{shimony, barnum, wei, wei2}, or more precisely, composite systems whose Hilbert space is the tensor product of that of its components.
The point is that, for spin systems, a generic wave function has no symmetry and one uses separable states to approximate it; while here for a system of identical fermions, the target wave function is always antisymmetric, and hence the simpler wave function used to approximate it should also be antisymmetric. We thus need to antisymmetrize a separable state and in doing so we get either a Slater determinant or zero.

We note that while the geometric entanglement of spin systems have been extensively studied (e.g., see \cite{wei, wei2, wei3, hayashi, hubener, aulbach, tamaryan10, orus08,  tamaryan08,   parashar, martin, chenpra, orus14} and references therein), it is much less discussed for identical particles, although some works on symmetric spin states \cite{hayashi,hubener} can be translated into the bosonic language too.  The reason might be that the extra symmetry or antisymmetry constraint makes the optimization problem seemingly more complicated \cite{benattipra, benattiJPB, benattiAP, kus}. Despite this difficulty, however, recently some progress has been made on the fermionic case \cite{zjm, zjm2,lin2015, lin3in6}. In particular, in \cite{zjm}, an efficient algorithm to construct the optimal Slater approximation of an arbitrary fermionic wave function was brought up. It is this algorithm that enables us to study the geometric entanglement in the Laughlin wave function.

Before proceeding to do so, we note that the advantage of the geometric entanglement to bipartite entanglements \cite{kaplan} is that, it is intrinsic, because it involves no partition, neither of the space nor of the particles. By avoiding partitioning the space (the modes), it is representation-independent; by avoiding partitioning the particles into two groups, it treats identical particles identically, i.e., on an equal footing. In contrast, the Schmidt decomposition based bipartite entanglement depends on the partition of the system, which is inevitably arbitrary.

\section{Geometric entanglement by the optimal Slater approximation}

The algorithm we will utilize was already detailed in \cite{zjm}. Here we sketch it briefly  for the sake of completeness, and tailor it somehow for the specific problem.

The Laughlin wave function of $N $ electrons, in the dimensionless form,  is ($z_i = x_i + i y_i $)
\begin{eqnarray}\label{lwf}
  \Psi &=& \mathcal{N}  \exp \left(- \frac{1}{4}\sum_{i=1}^N |z_i|^2 \right) \prod_{i< j } (z_i - z_j)^m .
\end{eqnarray}
Here $m$ is a positive odd integer and $\mathcal{N}$ is a normalization factor. Only the $m > 1$ states are of interest---the $m=1$ state is simply a Slater determinant \cite{yoshioka}. We shall focus on the $m=3$ and $m =5$ states. This is on the one hand due to the limitation of numerical resources and on the other hand because of the fact that the states with higher values of $m $ are less good as variational wave functions \cite{yoshioka}.

By construction, each particle in the Laughlin wave function (\ref{lwf}) is in the lowest Landau level. More specifically,
\begin{eqnarray}\label{lwf2}
   \Psi &=& \exp \left(- \frac{1}{4}\sum_{i=1}^N |z_i|^2 \right) \sum_{\alpha } A_\alpha z_1^{\alpha_1 -1} z_2^{\alpha_2 -1} \ldots z_N^{\alpha_N -1 } \nonumber \\
   &=& \sum_{\alpha } B_\alpha f_{\alpha_1} (z_1) f_{\alpha_2} (z_2) \ldots f_{\alpha_N} (z_N) .
\end{eqnarray}
Here $\alpha \equiv  (\alpha_1, \alpha_2, \ldots, \alpha_N )$ is an $N$-tuple with $1\leq \alpha_i \leq d = m (N-1) +1  $. In the second line, we have introduced the  single-particle orbitals ($n\geq 1 $)
\begin{eqnarray}\label{landauorbitals}
  f_n  (z ) &=& \frac{1}{\sqrt{2 \pi 2^{n-1 } ( n-1)! }} z^{n -1 } e^{-|z |^2/4} ,
\end{eqnarray}
which are an orthonormal basis of the lowest Landau level. By (\ref{lwf2}), we see that $\Psi $ is supported by the $d  $-dimensional single-particle Hilbert space of
\begin{eqnarray}
  \mathcal{H}_d =   Span \{ f_1, f_2, \ldots, f_{d} \}.
\end{eqnarray}
Actually, it is readily seen that these first $d$  $f$'s are also the natural orbitals \cite{lowdin} of $\Psi$ and each of them has a nonzero population. Because an $N$-particle Slater-determinant wave function has exactly $N$ populated natural orbitals (each of population 1), we see that the Laughlin wave function with $m>1$ is not a Slater determinant. The fact that $\Psi $ is supported by a finite dimensional single-particle Hilbert space greatly facilitates the numerical calculation.

The $N$-electron Hilbert space is spanned by the $ \left( \begin{array}[c]{c}d  \\N  \end{array} \right)$  Slater determinants
\begin{eqnarray}
  S_\beta  &=& \frac{1}{\sqrt{N! }}
  \left|
  \begin{array} [c]{ccc}%
                                                f_{\beta_1} (z_1) & \cdots & f_{\beta_1} (z_N) \\
                                                \vdots  & \ddots & \vdots  \\
                                                f_{\beta_N} (z_1) & \cdots  & f_{\beta_N} (z_N)
                                                \end{array}
                                              \right|,
\end{eqnarray}
with $\beta = (1\leq \beta_1< \beta_2< \ldots< \beta_N \leq d )$ being an $N$-tuple. The Laughlin wave function can be expanded in terms of these Slater determinants as
\begin{eqnarray}\label{expansion}
  \Psi  &=& \sum_\beta C_\beta S_\beta .
\end{eqnarray}
This expansion (to determine the coefficients $C_\beta $) is highly nontrivial. But fortunately,  some efficient algorithms, which we employ here, have been given in \cite{scharf}. By the way, we would like to mention that the dimension of the Hilbert space increases with $N$ exponentially. For $m=3$ and $N=10$, the dimension is $13\;123\;110$ and for $N=11$, it will be $84\;672\;315$. Because of the limitation of memory and time, in this work we have reached at most $N=10 $.

A generic Slater determinant is supported by an $N$-dimensional subspace $V $ of $\mathcal{H}_d$. Suppose $\{g_1, \ldots, g_N \}$ is an orthonormal basis of $V$. They are linearly related to the orthonormal basis $\{f_1, \ldots, f_d \}$ of $\mathcal{H}_d $ by a $d\times N $ matrix $M$, i.e.,
\begin{eqnarray}\label{gvsf}
  g_i  &=& \sum_{j=1}^{d} M_{ji  } f_j .
\end{eqnarray}
The condition that $\{g_i \}$ be orthonormal requires that
\begin{eqnarray}\label{MM}
M^\dagger M  = I_{N\times N } ,
\end{eqnarray}
the $N\times N $ identity matrix. The Slater determinant constructed out of $\{g_i \}$  is
\begin{eqnarray}
  S  &=& \frac{1}{\sqrt{N! }}
  \left|\begin{array}[c]{ccc}
                                                g_{1} (z_1) & \cdots & g_{1} (z_N) \\
                                                \vdots  & \ddots & \vdots  \\
                                                g_{N} (z_1) & \cdots  & g_{N} (z_N)
                                              \end{array}\right|.
\end{eqnarray}
Choosing another basis of $V$ will yield the same wave function up to a global phase.

The strategy to maximize the modulus of the overlap between $S $ and the Laughlin wave function $\Psi $ is very simple. We start from some  initial orbitals $\{g_i \}$ chosen randomly, and then fix all but one of them and try to optimize it with respect to the rest orbitals.  For example, let us fix orbitals $\{ g_{2\leq i \leq N }\}$ and try to find an optimal $g_1$ with respect to them. We note that the inner product between $S$ and $\Psi $ can be written as ($d \sigma_i \equiv  d x_i d y_i $)
\begin{eqnarray}\label{inner1}
  \langle \Psi | S \rangle  &=& \frac{1}{\sqrt{N!}} \int d \sigma_1 \ldots d \sigma_N \sum_{P} (-)^P g_{P_1} (z_1) \ldots g_{P_N}(z_N)    \Psi^* ( z_1, \ldots, z_N  ) \nonumber \\
  &=& \sqrt{N! } \int d \sigma_1 \ldots d \sigma_N
     g_{1} (z_1) \ldots g_{N}(z_N)  \Psi^* ( z_1, \ldots, z_N  ) \nonumber \\
     &=& \int d \sigma_1 g_1 (z_1) h^* (z_1) .
\end{eqnarray}
Here from the first line to the second line, we have used the fact that the wave function $ \Psi $ is antisymmetric and thus the $N! $ terms summing over $P $ are all equal actually. From the second line to the third line, the integrals over $z_{2\leq i \leq N }$ were performed, and we have defined a single-particle state
\begin{eqnarray}
  h(z_1 ) &=&  \sqrt{N! } \int d \sigma_2 \ldots d \sigma_N
     g_{2}^* (z_2) \ldots g_{N}^*(z_N)  \Psi (z_1, z_2, \ldots, z_N ).
\end{eqnarray}
Equation (\ref{inner1}) is in the form of the inner product of $g_1$ and $h$. By the Cauchy-Schwarz inequality, we immediately see that the optimal $g_1 $ should be proportional to $h $. Here an accidental and fortunate fact is that, $h$ is orthogonal to all $g_{2\leq i \leq N}$ because of the antisymmetry property of $\Psi $. Hence, when we update $g_1 $, the orthonormal condition (\ref{MM}) between the $g$'s  is still maintained.

\begin{figure*}[tb]
\centering
\includegraphics[width=0.46\textwidth]{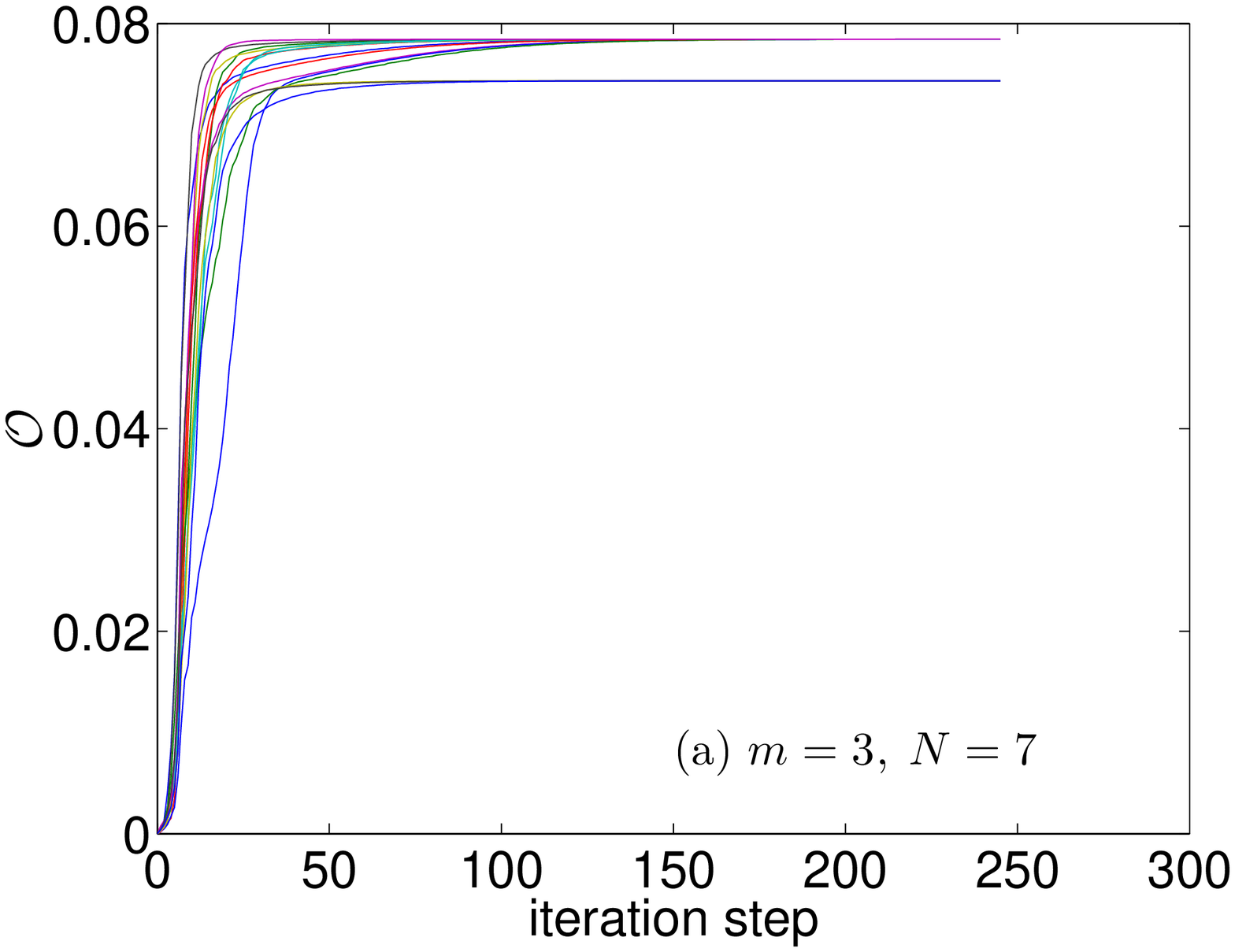}
\includegraphics[width=0.46\textwidth]{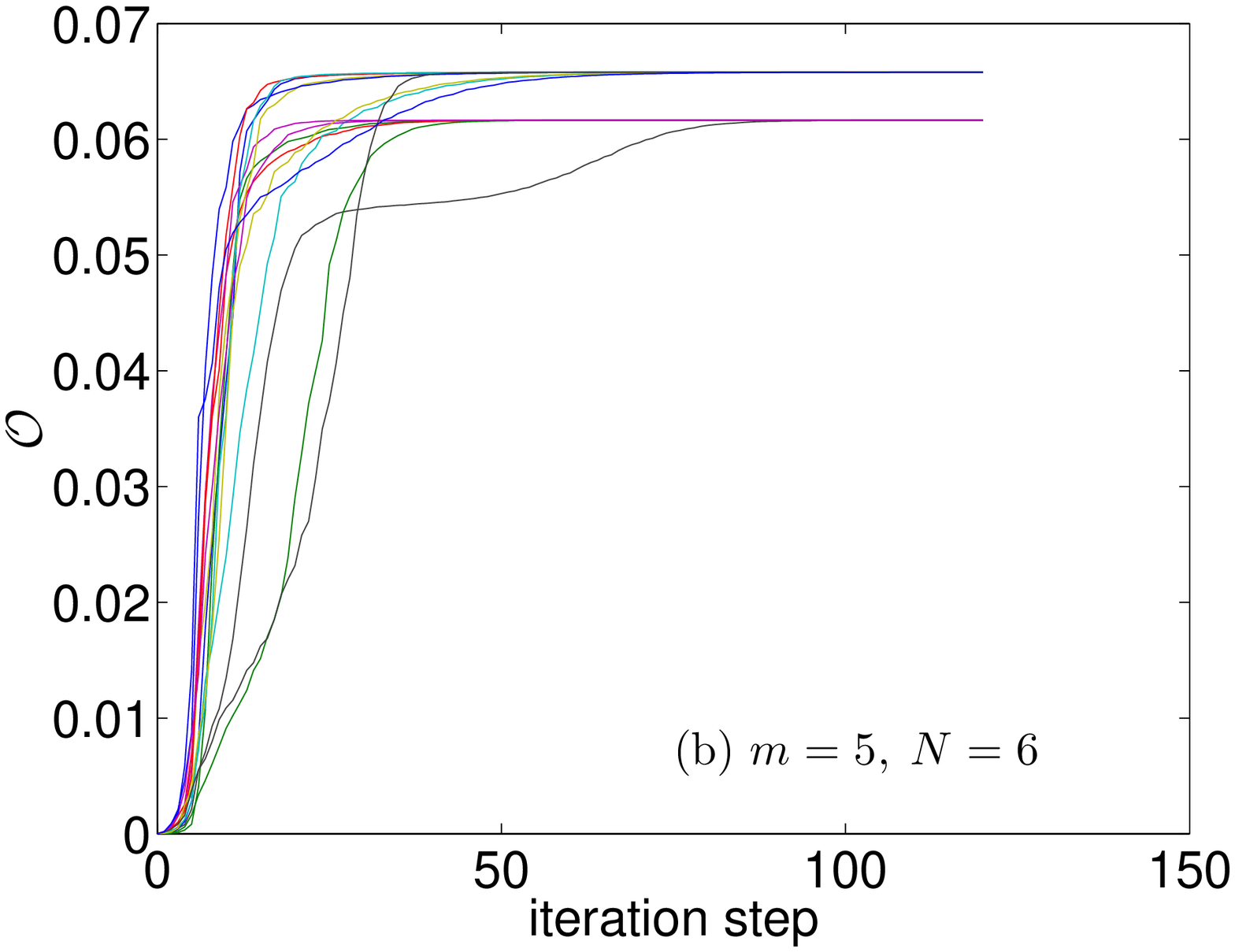}
\caption{(Color online) Evolving history of the overlap $\mathcal{O}$ for (a) $(m,N )= (3, 7)$ and (b) $(m,N) = (5,6)$. In each iteration step, one orbital is updated. In each panel, 15 trajectories, which correspond to 15 different initial guesses for the orbitals, are shown. Note that some trajectories evolve into local maxima (the lower plateaus). That is why running the program multiple times is necessary. }
\label{history}
\end{figure*}

The problem is then reduced to calculating $h$. Using (\ref{expansion}) and (\ref{gvsf}), it is straightforward to get the compact formula
\begin{eqnarray}
  h(z_1 ) &=& \sum_{\beta } C_\beta
   \left|\begin{array}[c]{cccc}
    f_{\beta_1} (z_1) & M_{\beta_1, 2}^* & \cdots & M_{\beta_1, N }^* \\
  \vdots & \vdots & \ddots & \vdots \\
  f_{\beta_N}(z_1) & M_{\beta_N, 2}^* & \cdots & M_{\beta_N, N}^*  \end{array}\right| .
\end{eqnarray}
We can then calculate $h $ using the Laplace expansion \cite{rose} for the determinant, and update the orbital $g_1$, or more specifically, update the first column of $M $. By doing so, the overlap $\mathcal{O}$ increases. Now the point is that, while $g_1$ has been chosen as the optimal one with respect to the rest $N-1$ orbitals, $g_2$ is not necessarily optimized with respect to its accompanying orbitals, namely $\{g_1, g_3, \ldots, g_N \}$. Therefore, we can turn to $g_2 $ and update it in a similar way. This procedure can be repeated by sweeping across the $N$ orbitals (the $N$ columns of $M $) in a circular way. The overlap increases after each update and surely will converge as it is upper bounded by unity. To make sure that we hit the global maximum instead of being trapped by some local one, the program has to be run dozens of times with different initial values of $M$. The whole process and the potential issues are illustrated in figure \ref{history}, where the evolution trajectories of the overlap $\mathcal{O}$ are shown. There we see that a significant portion of the trajectories settle down on the lower plateaus (local maxima), proving that it is absolutely necessary to start from a sufficiently large amount of sets of initial orbitals.

\begin{figure}[tb]
\centering
\includegraphics[width=0.45\textwidth]{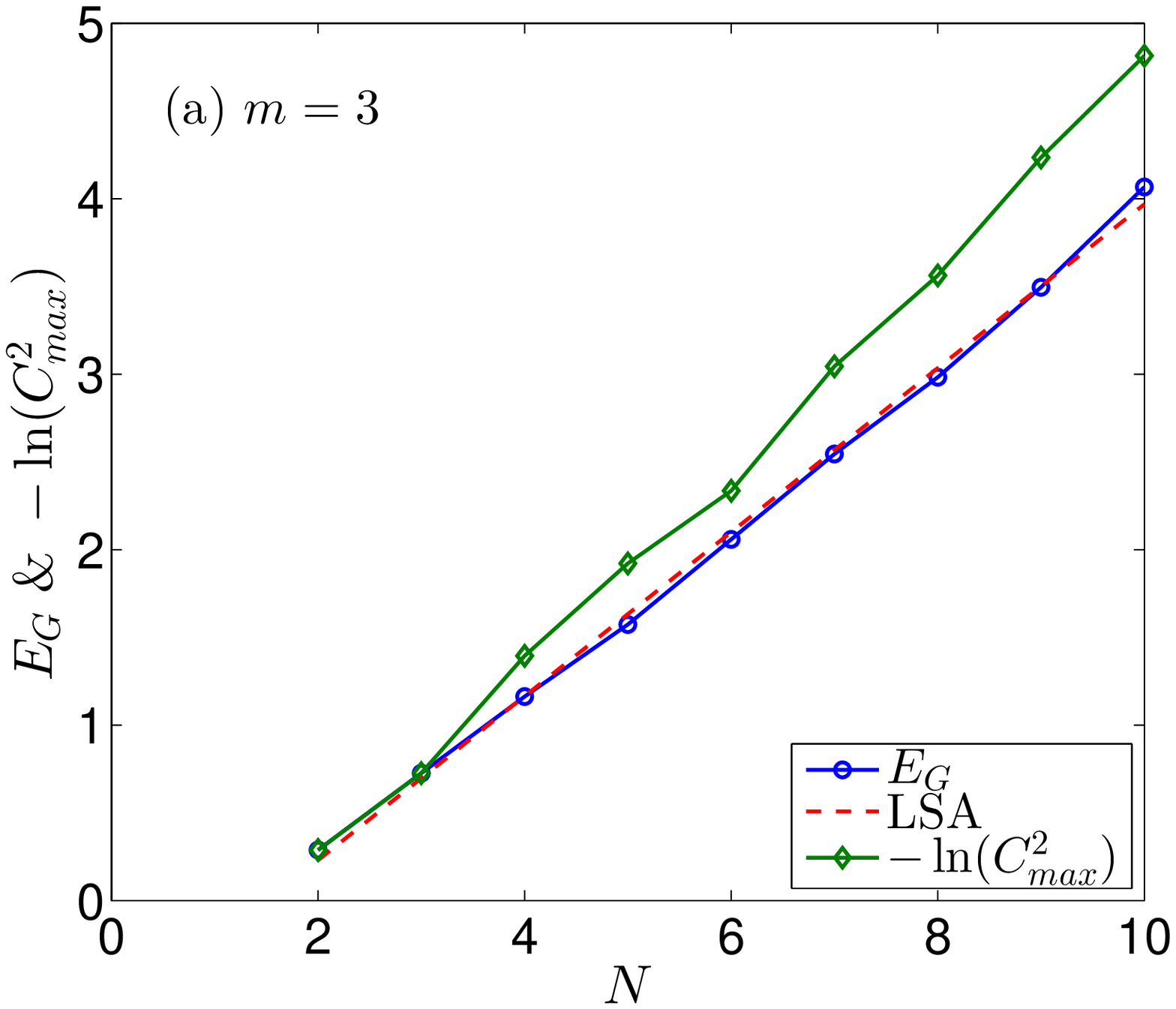}
\includegraphics[width=0.45\textwidth]{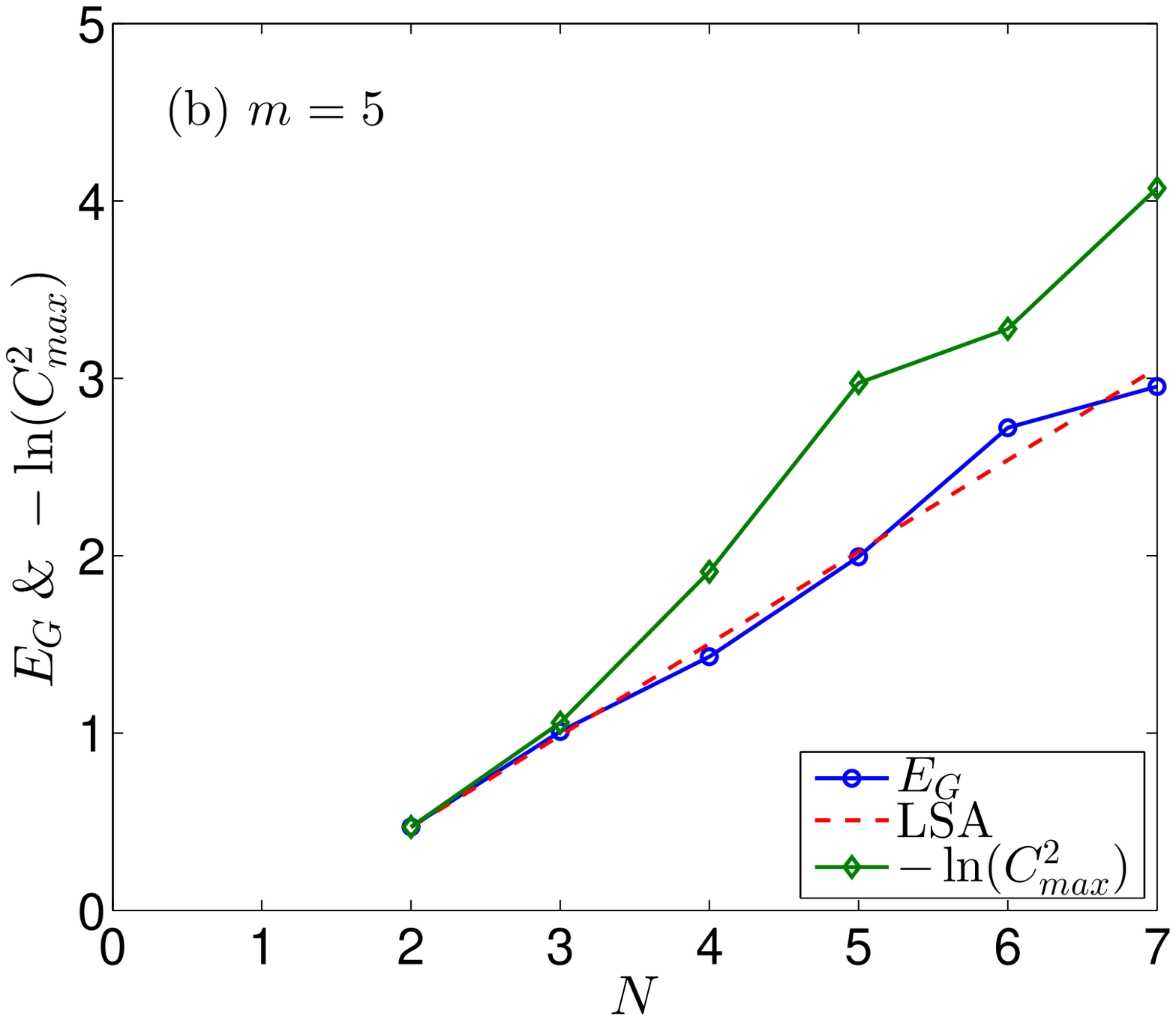}
\caption{(Color online) Geometric entanglement $ E_G $ (the $\circ  $ markers) between the electrons in the Laughlin wave function (\ref{lwf}), the least squares approximations (LSAs) to the exact data, and $-\ln (C_{max}^2)$ [the $\diamond $ markers, see (\ref{cmax})] as an upper bound of $E_G$.  }
\label{m3m5}
\end{figure}

In figure \ref{m3m5}, we present the geometric entanglement $E_G$ against the number of electrons, both for $m=3$ and $m=5$. Each data point is obtained with at least 30 different initial values of $M$, and the columns of $M$ are updated at least for 40 rounds. Out of the final values of $\mathcal{O}$, the maximal one is  taken as $\mathcal{O}_{\rm{max}} $, and then $E_G$ is calculated. Together with the exact value of $E_G$, also shown is its upper bound approximation $-\ln (C^2_{max})$, where
\begin{eqnarray}\label{cmax}
  C_{max} &\equiv & \max_{\beta } |C_{\beta}|  .
\end{eqnarray}
In view of (\ref{expansion}), $C^2_{max} $ is the maximal overlap between $\Psi$ and a Slater determinant constructed out of the Landau orbitals (\ref{landauorbitals}). In the two-particle case of $N=2$, we have $E_G = - \ln (C^2_{max}) $ because coincidentally the expansion (\ref{expansion}) is the canonical expansion of the two-fermion wave function $\Psi$, and in this case we have the theorem that the optimal Slater approximation is exactly the most weighted term in the expansion \cite{zjm}. However, for $N> 2$, we see that generally $E_G $ is significantly smaller than $- \ln (C^2_{max}) $, which means that by \emph{mixing} the Landau orbitals appropriately, the overlap between the Slater determinant and the Laughlin wave function can be increased.

We note that the $m=3$ case is surprisingly regular. First, all the points lie very close to a straight line. This linear behavior is not obvious at all, as we know that for the maximally entangled Dicke state, in which half of the spins are up and half are down,  $E_G $ is not linear in the number of spins but logarithm \cite{wei3, hayashi}. Second, even if $E_G $ is found to be linear asymptotically in the large $N $ limit;  in the small $N $ limit, deviation from this behavior is expected.  That is, there should be some crossover regime. However, the linear behavior extends well down to the smallest possible value of $N =2$. The $m=5$ case is less regular. However, again there is no apparent crossover regime at the lower limit of $ N = 2$.

The pattern of the data points suggests using the linear least squares approximation to fit the data. That is,
\begin{eqnarray}
  E_G &\simeq & a  N - \gamma .
\end{eqnarray}
We get for $m=3$,
\begin{eqnarray}\label{m3}
  \gamma = 0.7038 \pm 0.0949
\end{eqnarray}
and for $m=5$,
\begin{eqnarray}\label{m5}
   \gamma = 0.5662  \pm 0.2537 .
\end{eqnarray}
In \cite{orus14}, it is found that for some topologically ordered systems such as the toric code, double semion, color code, and quantum double models, the geometric entanglement, like the bipartite entanglement based on space partition, consists of a  term proportional to $N$ plus a constant term called the topological entropy. If this is also the case for the Laughlin wave function, we would have the constant term $\gamma $ as $\ln \sqrt{3} = 0.5493 $ for $m=3$, and $\ln \sqrt{5}  = 0.8047 $ for $m = 5$. However, this is not the case as shown by (\ref{m3}) and (\ref{m5}). Actually, even the trend is in the wrong direction.

\begin{figure}[tb]
\centering
\includegraphics[width=0.45\textwidth]{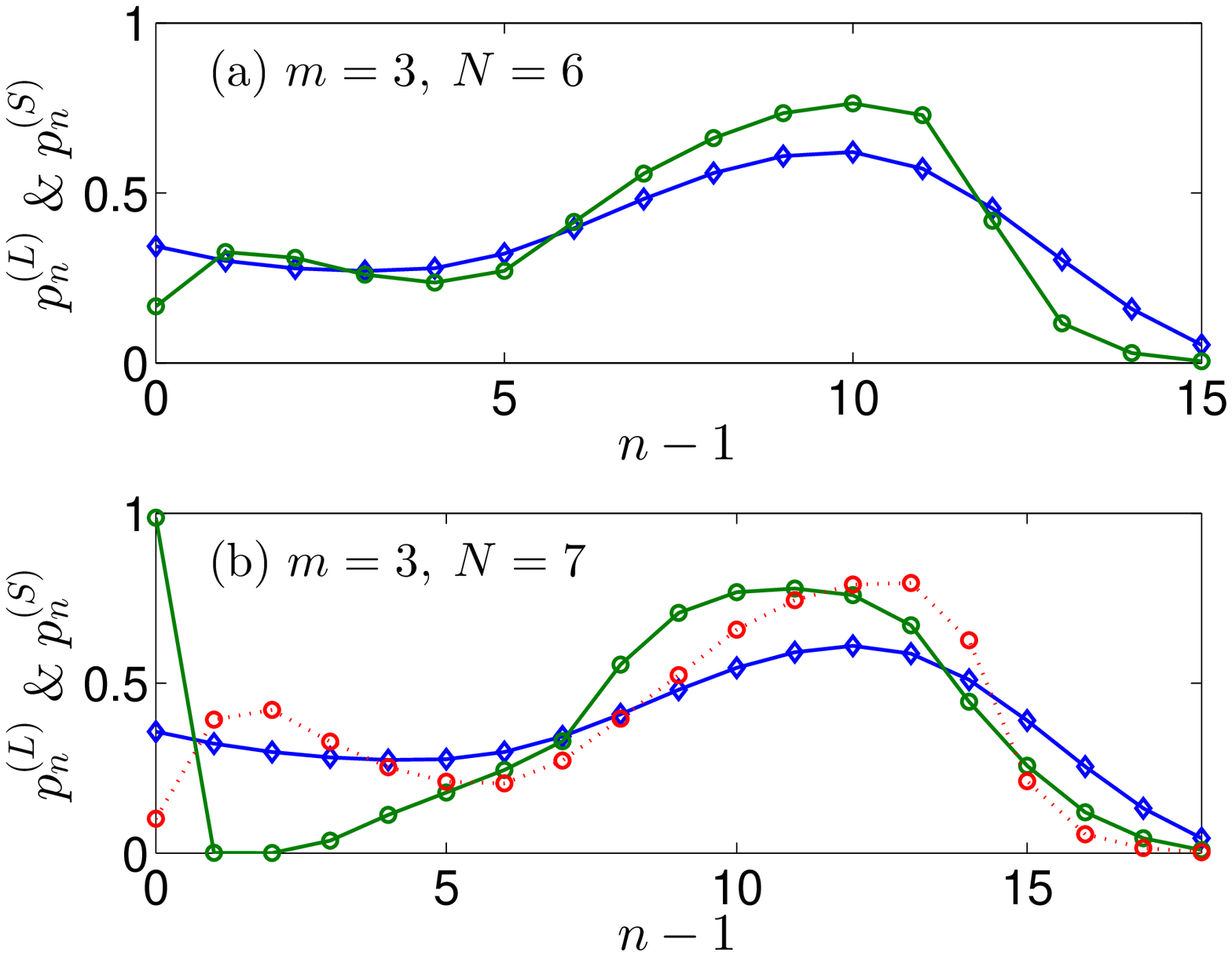}
\includegraphics[width=0.45\textwidth]{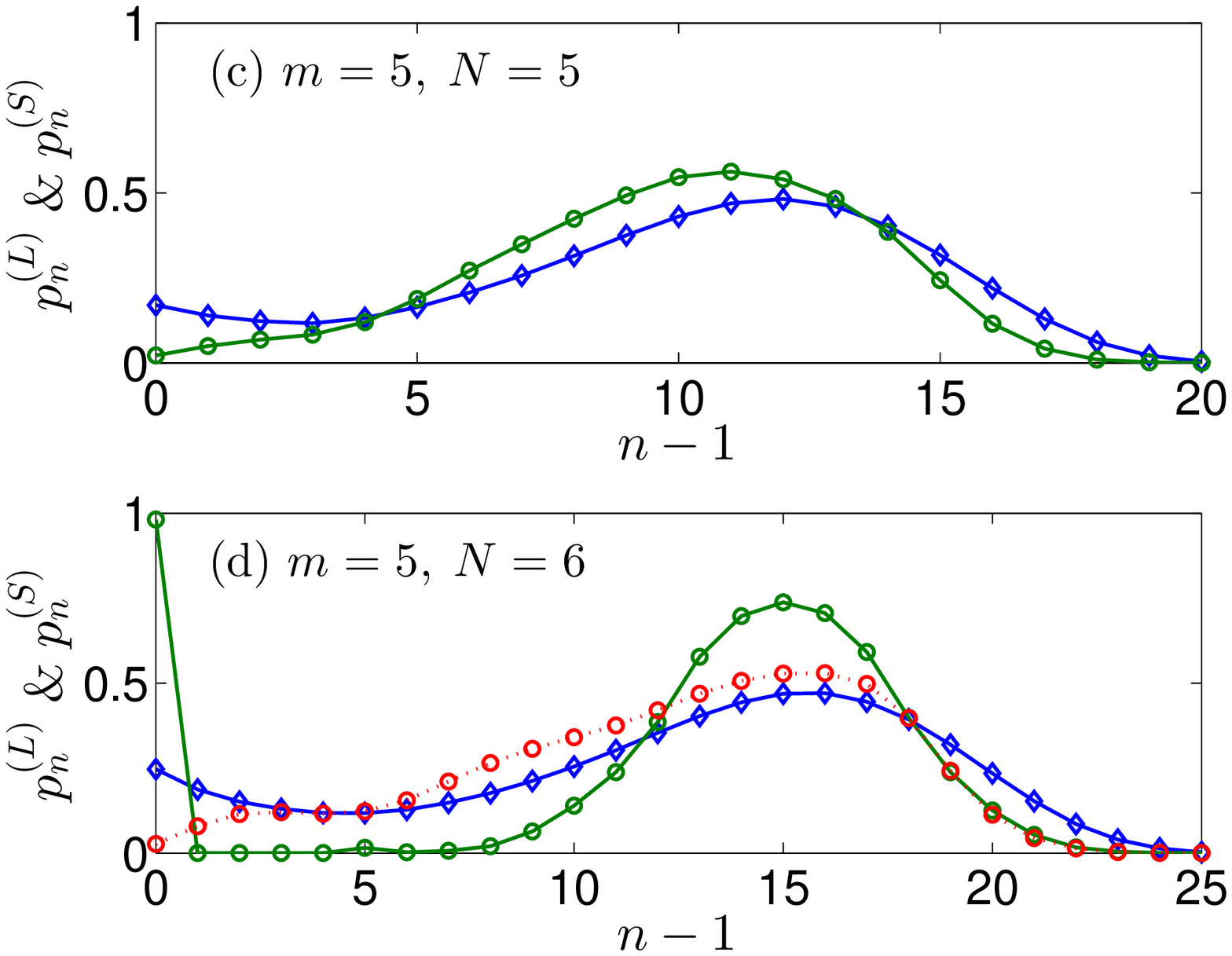}
\caption{(Color online) Populations on the Landau orbitals (\ref{landauorbitals}). The $\diamond$ markers correspond to the Laughlin wave function [see (\ref{pnl})]. The $\circ$ markers chained with solid lines correspond to the optimal Slater approximation [see (\ref{pns})]. In (b) and (d), the $\circ$ markers chained with dotted lines belong to the less optimal Slater determinant which corresponds to the lower plateaus in figure \ref{history}.  }
\label{figbest}
\end{figure}

When we get the maximal overlap $\mathcal{O}_{max}$, we get simultaneously the $N $ single-particle orbitals $\{g_i | 1\leq i \leq N \}$, which are encoded in the columns of the matrix $M $. It is thus possible and instructive to have a look of these orbitals and the corresponding Slater determinant. As emphasized above, the orbitals are not invariant; what is invariant is the $N$-dimensional subspace spanned by them. In other words, the $d\times N $ matrix $M $ is not invariant, but the $d\times d $ matrix $M M^\dagger$ is invariant. The latter quantity is actually the one-particle reduced density matrix (in the representation of the Landau orbitals) of the Slater determinant constructed out of the $g$'s. A  quantity of particular interest is the population of the optimal Slater determinant on the $n$th Landau orbital, which is the $n$th diagonal element of $M M^\dagger $, namely ($1\leq n \leq d $)
\begin{eqnarray}\label{pns}
  p_n^{(S)} &=&  \sum_{i=1}^N |\langle f_n | g_i \rangle |^2 = \sum_{i=1}^N |M_{ni }|^2 .
\end{eqnarray}
The same quantity characterizing the Laughlin wave function $\Psi $ is
\begin{eqnarray}\label{pnl}
  p_n^{(L)} &=& \sum_{\beta \ni n } |C_\beta |^2.
\end{eqnarray}
In figure \ref{figbest}, $p_n^{(S)}$ and $p_n^{(L)}$ are shown for some pairs of $(m,N)$. We see that in figures \ref{figbest}(a) and \ref{figbest}(c), $p_n^{(S)}$ and $p_n^{(L)}$ are somewhat close. However, as $N$ increases by 1, as in figures \ref{figbest}(b) and (d), the shape of the curves of $p_n^{(S)}$ changes abruptly. In both figure \ref{figbest}(b) and \ref{figbest}(d), $p_1^{(S)}$ is very close to unity (but not exactly), while $p_{2\leq n\leq 3}^{(S)}$ is very close to zero in \ref{figbest}(b) and $p_{2\leq n\leq 5}^{(S)}$ is very close to zero (less than $10^{-4}$) in \ref{figbest}(d). This feature is then preserved for higher value of $N$ for both values of $m$. The latter fact is compatible with the former. As can be seen from the expression of the Laughlin wave function (\ref{lwf}), if some electron is in the $n=1$ landau orbital, then all other electrons will be at least in the $n=m+1$ orbital. Hence, if in the optimal Slater determinant the $n=1$ landau orbital is almost completely occupied, then it is advantageous to avoid the $2\leq n \leq m $ orbitals. We notice that figures \ref{figbest}(b) and (d) are of the same parameters as figures \ref{history}(a) and (b), which exhibit competing local maxima of ${ \mathcal O }$. We are thus led to study the suboptimal Slater determinants too and the associated orbitals. The corresponding occupation numbers ${p}_{n}^{(S)}$ are shown with the dotted lines in figures \ref{figbest}(b) and (d). It turns out that the curves are similar to those in \ref{figbest}(a) and (c). Hence, we see that there is a subtle competition between different configurations of the optimal orbitals in the problem in question---in \ref{figbest}(b) and (d), the configurations similar to the optimal configurations in \ref{figbest}(a) and (c) are overtaken by some drastically different ones.

\section{Conclusions and Discussions}

In conclusion, we have used an iterative algorithm to find the optimal Slater approximation of the Laughlin wave function and in turn to calculate the geometric entanglement in it. To the best of our knowledge, this is the first time that the geometric entanglement in a strongly interacting fermionic system, which is of great physical interest, is calculated. The importance of the Laughlin wave function warrants such a study.

We find that the geometric entanglement $E_G $ scales linearly with the electron number $N $. While this is not so surprising, it is unexpected that the linear behavior extends well down to the lower limit of $N = 2$. At the time being, a full understanding of this finding  is still lacking. But, it is in alignment with the fact that the Laughlin wave function has a very short correlation length (it is on the order of the magnetic length; or unity by our notation) \cite{yoshioka}, which means that the $N = 2$ case is already close to the thermodynamic limit. Actually, for the fractional quantum Hall system, many quantities behave similarly. For example, by exact diagonalization, it is found that in the $1/3$ filling case, the ground-state energy per particle for a four-electron system is already very close to the infinite system result \cite{Chakraborty}.

The linear behavior of the geometric entanglement prompted us to compare the constant term with the expected topological entropy. As it turns out, they do not agree. This is in contrast to \cite{orus14}, where the topological entropy can be successfully extracted from the scaling behavior of the geometric entanglement. Our numerical result as a negative case then indicates that the relation between geometric entanglement and topological entropy is subtle. Actually, we note that for indistinguishable particles, the concepts of geometric entanglement and topological entropy belong to two incompatible frameworks. The former allows no bipartition, neither of the particles nor of the space, while the latter pertains to the practice of bipartitioning the system. From this point of view, it is normal that we have failed to extract the expected topological entropy from the geometric entanglement---The former is more likely to be extracted from a bipartition-based measure of entanglement, as is done in \cite{masud}, where the orbital entanglement is calculated. As for \cite{orus14}, they have the advantage that they are dealing with spin systems, which enjoys the tensor product structure that we lack. Because of this, geometric entanglement and spatial partition are compatible. But even so, there exists the issue that for geometric entanglement, the system should be (and is) partitioned into multiple pieces instead of two. In view of this, there might be some deep reasons behind their success, if it is not a fortuitous coincidence.


%


\section*{Acknowledgments}

We are grateful to K. Yang, F. Pollman, and T.-C. Wei for stimulating discussions. This work is supported by the Fujian Provincial Science Foundation under Grant No.~2016J05004, NSF of China  under Grant No.~11504061, the China Postdoctoral Science Foundation,  and the Foundation of LCP.



\begin{thebibliography}{99}
\bibitem{nobel}
As a matter of fact, it is one of the two variational wave functions winning a Nobel prize. The other one is the BCS wave function.
\bibitem{interaction}
Note that we are talking about the generic case. Interaction does not necessarily mean the total wave function cannot be written as a Slater determinant. For example, for $N$ fermions in $N+1$ orbitals, the wave function is always a Slater determinant, although superficially it is the linear superposition of $N+1$ Slater determinants. See Appendix A in \cite{zjm} or Lemma 1 in \cite{zjm2} for more details.
\end{thebibliography}
\end{document}